\documentclass[%
 reprint,
superscriptaddress,
 amsmath,amssymb,
 aps,
 twocolumn,
longbibliography,
]{revtex4}

\usepackage{graphicx}
\usepackage{dcolumn}
\usepackage{bm}
\usepackage{nowidow}
\usepackage{tabularx}
\usepackage{siunitx}
\usepackage[dvipsnames]{xcolor}

\usepackage{xcolor}

\begin{document}

\newcommand{\mum}[1]{${#1}$~\textmu m} 

\title{Stress accumulation around ice in a temperature gradient}

\author{Dominic Gerber}
\affiliation{Department of Materials, ETH Z\"{u}rich, 8093 Z\"{u}rich, Switzerland.}%

\author{Lawrence A. Wilen}
\affiliation{Center for Engineering Innovation and Design, School of Engineering and Applied Sciences, Yale University, New Haven, Connecticut 06520, USA}%

\author{\mbox{Florian Poydenot}}
\affiliation{Laboratoire de Physique de l’\'{E}cole Normale Sup\'{e}rieure, UMR8023, CNRS, Universit\'{e} de Paris, PSL Research University, 75005 Paris, France}%

\author{Eric R. Dufresne}
\affiliation{Department of Materials, ETH Z\"{u}rich, 8093 Z\"{u}rich, Switzerland.}%

\author{Robert W. Style}
\affiliation{Department of Materials, ETH Z\"{u}rich, 8093 Z\"{u}rich, Switzerland.}%

\begin{abstract}
When materials freeze, they often undergo damage due to ice growth.
Although this damage is commonly ascribed to the volumetric expansion of water upon freezing, it is usually driven by suction of water towards growing ice crystals.
The freezing of this additional water can cause a large build up of stress. 
Here, we study this process by producing a  stable ice/water interface in a controlled temperature gradient, and measuring the deformation of the confining boundary.  
Analysis of the deformation field reveals  stresses applied to the boundary with $\mathcal{O}(\mu\mathrm{m})$ resolution.
Globally, stresses increase steadily over time as liquid water is transported to more deeply undercooled regions.  
Locally, stresses increase until ice growth is stalled by the confining stresses.  
In accordance with the Clapeyron equation, the local limiting stress is proportional to the local undercooling. 
These results are closely connected to the crystallization pressure for growing crystals
and 
condensation pressure during liquid-liquid phase separation.
\end{abstract}

\maketitle

We all know that when water freezes, its volume increases by a few percent.
While this explains how a filled water bottle  cracks in a freezer, it does not explain the huge deformations underlying freezing damage to roads (\emph{i.e.} frost heave), food, or biological tissue ~\cite{dash_physics_2006}. 
These more dramatic effects typically occur in a temperature gradient, and are associated with a flux of liquid water into regions below freezing, called \emph{cryosuction}.

\begin{figure}
    \centering
    \includegraphics[width=\linewidth]{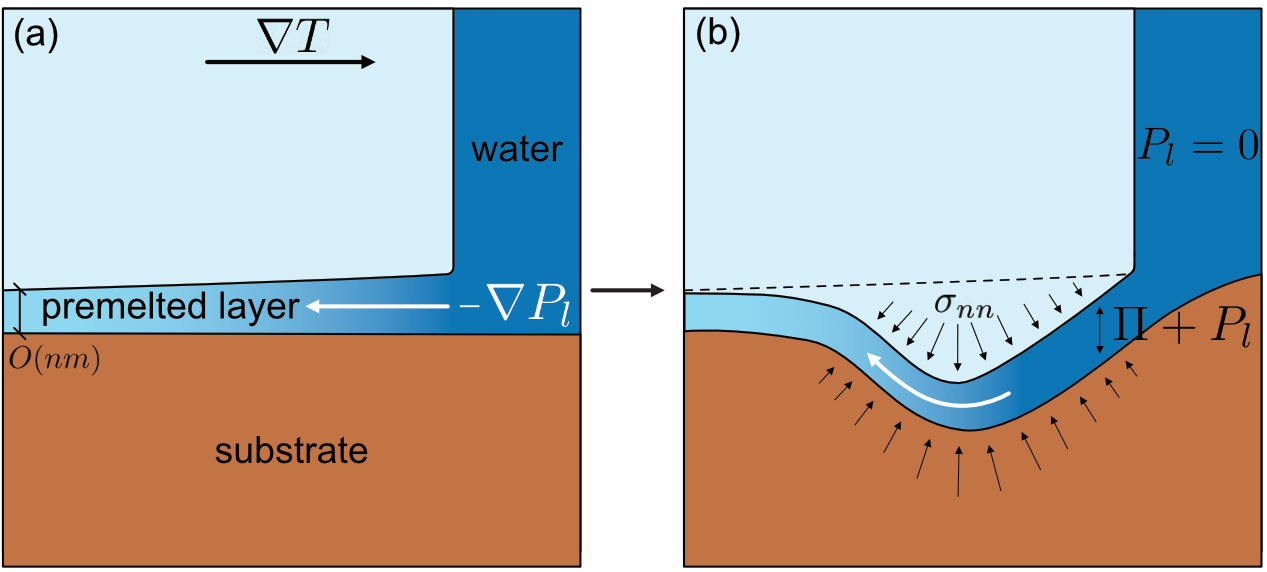} 
    \caption{Schematic showing how stresses build up due to cryosuction in a temperature gradient. (a) initially both ice and substrate are stress-free. (b) cryosuction causes ice accretion and stress build up near the bulk  ice/water interface.}
    \label{fig:schematic}
\end{figure}

Transport of liquid water into frozen regions occurs at boundaries between ice and its confining material.
Here, interfacial forces combine to give a repulsive interaction between the ice and the surrounding substrate, which result in a thin, mobile layer of liquid water between the ice and surrounding material (c.f. Fig. \ref{fig:schematic}, \cite{style_surface_2005,wilen_dispersionforce_1995,sibley_how_2021,wilen_dispersionforce_1995,sadtchenko_interfacial_2002,furukawa_ellipsometric_1987,murata_thermodynamic_2016,tyagi2020objects}).
The combined effect of the interfacial forces is given by a disjoining pressure, $\Pi=-\sigma_{nn} - P_l$, with $P_l$ being the effective pressure in the film, and $\sigma_{nn}$ being the normal stress exerted by the ice on the substrate.
Liquid flow is driven by gradients in $P_l$.
To calculate this, we note that $\Pi$ is predicted to be just a function of the local temperature.
Assuming local thermodynamic equilibrium between ice and water, and that the stress field in the ice is purely isotropic, yields the Clapeyron equation \cite{dash_physics_2006,wettlaufer_premelting_2006,black_applications_1995},
\begin{equation}
   \Pi =  -\sigma_{nn} – P_l=\rho q_m\frac{T_m-T}{T_m}.
   \label{eqn:clausius}
\end{equation}
Here, $\rho$ is the ice density, $T_m$ is the bulk freezing temperature at atmospheric pressure, and $q_m$ is the latent heat of melting.
This simplified form further ignores the density difference between water and ice, and the anisotropy of the crystal. 
With this, we obtain the useful prediction that flow can arise only due to gradients in $\sigma_{nn}$ or $T$. 

An example demonstrating how flow occurs in premelted films is shown in Figure~\ref{fig:schematic}.
Here, ice is formed in a static temperature gradient, $\nabla T$.
Assuming that the ice is initially stress-free, Eq. \ref{eqn:clausius} implies that $\nabla P_l\sim~\nabla T$, causing a  flow into and along the premelted film, from hot to cold.
As the water moves to colder temperature it freezes onto the ice, leading to growth of ice, as shown in Figure~\ref{fig:schematic}b.
Growth in this direction deforms the confining boundary and  creates stresses that can ultimate damage it.

These  theoretical concepts  form the basis of the majority of theories of cryosuction~\cite{wettlaufer_theory_1996, wettlaufer_premelting_2006,vlahou_freeze_2015,wettlaufer_dynamics_1995,gagliardi_nonequilibrium_2019} and frost heaving~\cite{gilpin_model_1980, derjaguin_flow_1986, rempel_microscopic_2011, peppin_physics_2013,style_kinetics_2012,konrad_mechanistic_1980}. 
However, they have, almost exclusively been tested via macroscale experiments which measure spatially averaged stresses and ice segregation in porous materials without microscale resolution, or without temperature gradients 
 \cite{ketcham_frost_1997,schollick_segregated_2016,beaudoin_mechanism_1974,you_situ_2018,watanabe_amount_2002,ozawa_segregated_1989,konrad_influence_1988,zhou_ice_2020,vignes_model_1974}.

Here, we directly measure the evolving stresses exerted by individual ice crystals in a steady temperature gradient with unprecedented spatial resolution.
We grow ice into a soft-walled chamber \cite{wilen_frost_1995}, and monitor the wall deformations, allowing us to measure evolving ice stresses.
Locally, stresses increase until they stall at a fixed value.
Globally, stresses continuously accumulate over increasingly large areas of the ice/substrate interface.

\begin{figure}
    \centering
    \includegraphics{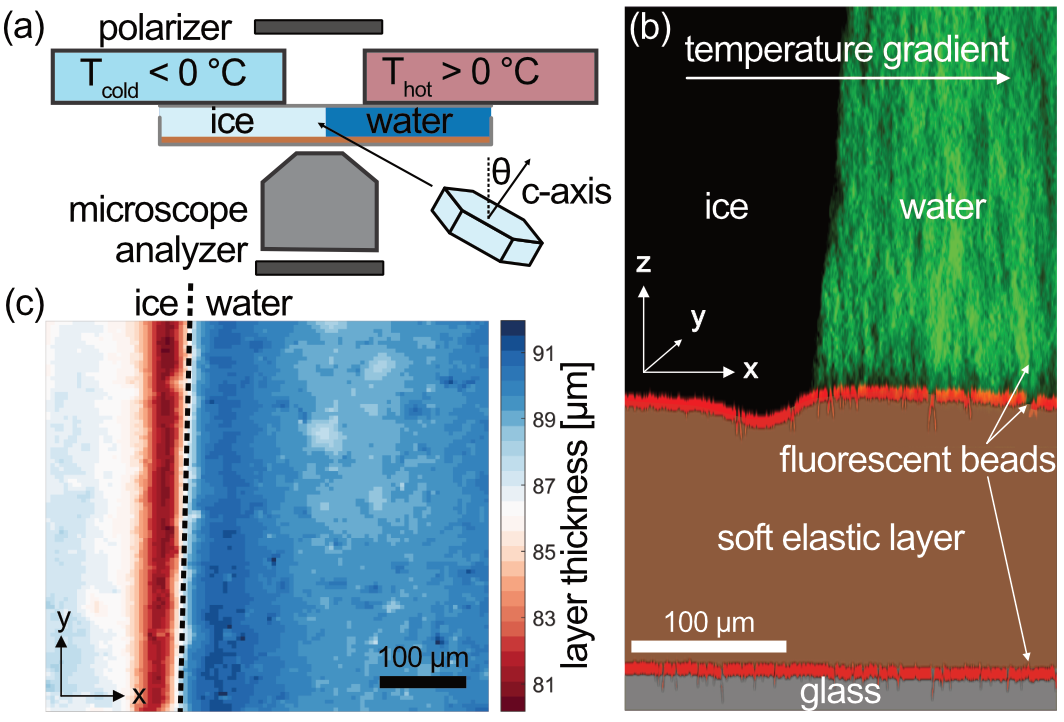} 
    \caption{Imaging substrate deformations in a freezing cell (a) A schematic of the experimental setup (b) Pseudo-colored confocal microscopy image showing a side view of the ice/water interface and the growing indentation into the soft elastic layer below it. The color in the water comes from fluorescent nanoparticles which are expelled by the growing ice. (c) A typical surface profile underneath a steady ice crystal. The profile is uniform along the ice/water interface, allowing us to collapse the data along the $y$-axis.}
    \label{fig:setup}
\end{figure}

Our experimental freezing setup, an extension of  \cite{dedovets_temperaturecontrolled_2018}, is shown schematically in Figure~\ref{fig:setup}a.
The freezing cell consists of two glass coverslips, with a \mum{600} spacing, where the  bottom coverslip is coated with a thin layer of silicone \cite{style_liquid-liquid_2018,kim_extreme_2020}.
The layer thickness is adjusted in the range of \mum{88-105}, and Young's modulus, $E=22-295$~kPa (characterized following \cite{garcia_determination_2018}).
To visualize deformations, 200 nm-diameter fluorescent tracer particles are attached to the silicone gel surface, following the protocol in  \cite{style_traction_2014}, with an average spacing of \mum{10}.
The freezing cell is mounted on the bottom of two aluminium blocks, whose temperature is fixed with a precision of $\pm$0.05~$^\circ$C, allowing us to impose a steady temperature gradient across the cell.
With this setup, the ice/water interface position fluctuates by less than \mum{5} in experiments lasting up to 6~hours.
The exact value of $\nabla T$ is measured by placing a thermistor in the cell and measuring its distance to the ice/water interface.
A 2~mm gap between the blocks lets us image ice growth with brightfield, polarized light, and confocal microscopy \cite{hell_aberrations_1993,besseling_methods_2015}.
See the Supplementary information for further details.

To perform the experiment, we fill a cell with de-ionized water that has been allowed to equilibrate with the room atmosphere (i.e. it contains a small amount of dissolved air \cite{bauer_chemical_1980}).
We also add a minimal amount of fluorescent, 100nm-diameter particles to visualize the ice/water interface (Fig.~\ref{fig:setup}b). 
The cell is cooled to $-0.5^\circ$C on one side, and ice is nucleated by touching it briefly with a liquid-nitrogen-soaked cotton swab.
The temperature is slowly reduced (at 0.1~K/min) to grow ice to the edge of the intended imaging region.
Then, after 15 minutes of equilibration time, we start taking confocal stacks and brightfield images.
While imaging, we  apply one last small cooling step, where
 we advance the ice into an initially stress-free region.
In this way, we  are able to capture the very beginning of the stress build-up.
The process creates large individual crystals, so that our entire imaging region consists of a single ice crystal.
We measure its orientation by looking at the crystal color between crossed polarizers (\cite{bloss_optical_1999,m_bauer_githubcalculated_michel_levy_chart_2019,sorensen_revised_2013,wilen_development_2003} see Supplementary information). 
This allows us to measure the angle, $\theta$, between the $c$-axis of the ice crystal and the $z$-axis with about $10^\circ$ accuracy.
If the ice exerts stresses on the surrounding cell, we see this as deformations of the silicone layer.
We quantify this by measuring the positions of the individual tracer particles with a confocal microscope, and tracking their displacements with a resolution of \mum{0.06} in the $x$, $y$ directions and \mum{0.4} in the $z$ direction, with $x,y,z$ defined in Figure \ref{fig:setup}b.

Since the cell is open to atmospheric pressure at both ends, there is no pressure build up due to water expanding as it freezes.
However, stresses still build up next to the ice/water interface due to cryosuction.
This can be seen in the raw microscopy data in Figure~\ref{fig:setup}c where there is a clear indentation in the soft elastic layer next to the ice/water interface. 
For single crystals, this deformation is always very uniform along the ice/water interface, as shown in Figure~\ref{fig:setup}c, and thus we average it along the axis of the ice/water/substrate contact line to give 2D representations of the substrate indentations, as shown in Figure \ref{fig:setup}b and subsequent figures.

\begin{figure*}
    \centering
    \includegraphics[width=0.9\linewidth]{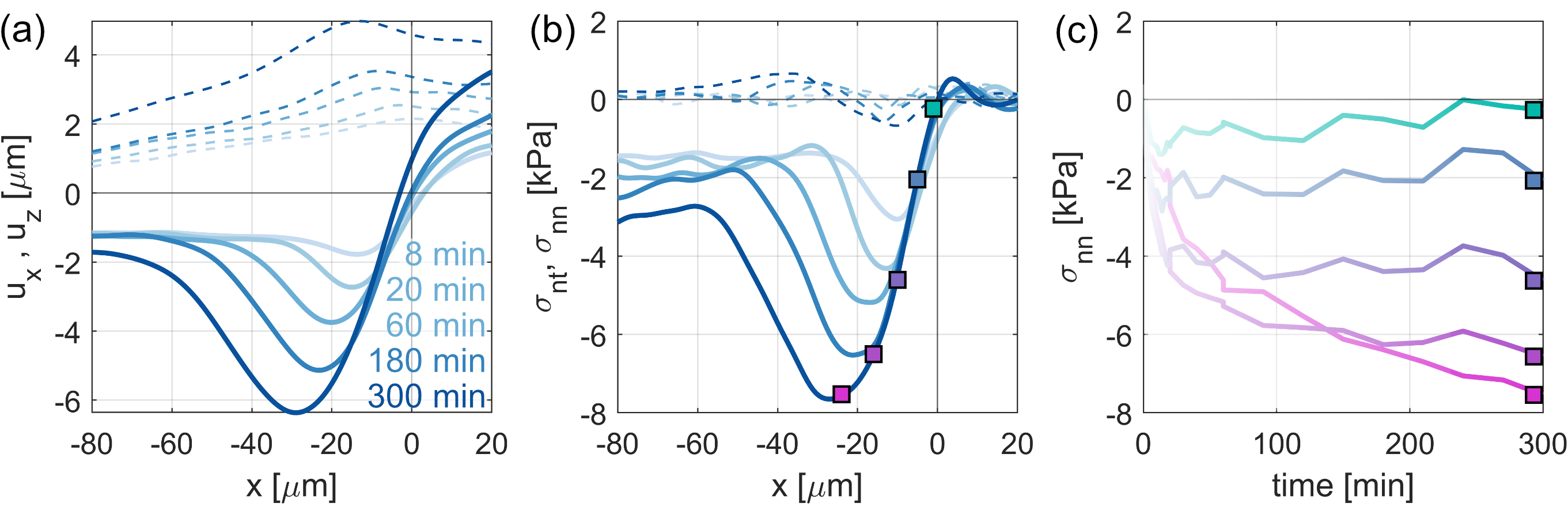}
    \caption{The time evolution of surface displacements and stresses near an ice/water interface at $x=0$ (with $E=22$~kPa, $G=0.4$~K/mm and thickness of \mum{88}) for (a) Substrate displacements $u_x$ (dashed curves) and $u_z$ (continuous curves) (b) The corresponding stresses, $\sigma_{nt}$ (dashed curves) and $\sigma_{nn}$ (continuous curves) (c) Evolution of $\sigma_{nn}$ over time at the positions marked in panel (b).}
    \label{fig:deform_and_stresses_over_time}
\end{figure*}

A typical example of the evolution of substrate deformations under the ice is shown in Figure~\ref{fig:deform_and_stresses_over_time}a.
This shows measured horizontal and vertical displacements of the surface ($u_x$ and $u_z$ respectively) at selected timepoints.
At each timepoint, these are plotted relative to the position of the ice/water contact line on the substrate.
At time zero, the final cooling step is applied to advance the ice.
The ice takes approximately 10 minutes to stabilize at its final position, and there is a small amount of substrate deformation during this time.
However, after this, we see much larger deformations appear with a continuously growing dimple emerging behind the ice front.
Simultaneously the substrate bulges up on the liquid side of the contact line, due to its incompressibility.
Initially the dimple grows quickly, but later slows down significantly (note the growing time intervals).
As the indentation grows, its position of maximum indentation also moves backward, resulting in a significant volume of ice pushing down into the underlying substrate.

For further insight into these deformations, we calculate the stresses exerted by the ice with Traction Force Microscopy (TFM)~\cite{style_traction_2014,sabass_high_2008}.
This involves solving a linear-elastic problem to calculate the in- and out-of-plane traction stresses exerted on the silicone surface ($\sigma_{xz}$, $\sigma_{zz}$ respectively) from $u_x$ and $u_z$ \cite{xu_imaging_2010}.
In brief, point values of $u_x$ and $u_z$ are interpolated onto a grid and smoothed to reduce measurement noise.
These fields are Fourier-transformed and multiplied by a kernel function, $Q(E,\nu,h)$, where $\nu$ and $h$ are the substrate's Poisson ratio and thickness respectively.
We obtain the traction stresses by inverse Fourier transforming and subtracting a constant value from $\sigma_{zz}$ to ensure that the average traction stresses on the water side of the interface are zero.
The latter step is necessary to give absolute values of $\sigma_{zz}$ when using incompressible substrates, but is not required for calculating $\sigma_{xz}$ \cite{xu_imaging_2010}.
Finally, we rotate these stresses to get $\sigma_{nn}$ and $\sigma_{nt}$, the normal and shear stresses relative to the local substrate surface.
$\sigma_{zz}=\sigma_{nn}$ to excellent approximation, but there is a small quantitative difference between $\sigma_{nt}$ and $\sigma_{xz}$.
Further details are given in the Supplementary Information.

The TFM results show that the stresses build up near the contact line, similarly to the substrate deformations (Figure~\ref{fig:deform_and_stresses_over_time}b).
However, unlike the indentation, both sets of stresses under the water-filled part of the cell are flat and close to zero.
This makes sense because the cell is open to the atmosphere and there is no pressure in the liquid phase.
Furthermore the traction stresses are approximately an order of magnitude smaller than the normal stresses, and the measured horizontal deformations $u_x$ are predominantly caused by the normal stresses.
This fits with the schematic picture in Figure \ref{fig:schematic} of a lubricated interface between the silicone and the ice.

While the global maximum stress steadily increases, the stress at each position appears to eventually saturate.
In Figure~\ref{fig:deform_and_stresses_over_time}c, we plot $\sigma_{nn}$ versus time at fixed positions behind the ice/water interface.
We see that the stresses close to the ice/water interface  plateau, suggesting the presence of a local maximum pressure that the ice can exert on its surroundings.
The further away from the interface, the longer it takes for the stress to stall.  At the furthest point from the interface shown (pink curve in Figure~\ref{fig:deform_and_stresses_over_time}c),  the stress build-up slows, but doesn't plateau over the course of the experiment.
For points close to the interface there appears to be a small but steady increase in the plateau region. 
We believe this is caused by a small drift of the ice/water interface position over the course of the experiment (see Supplementary Information).

\begin{figure*}
    \centering
    \includegraphics[width=1\linewidth]{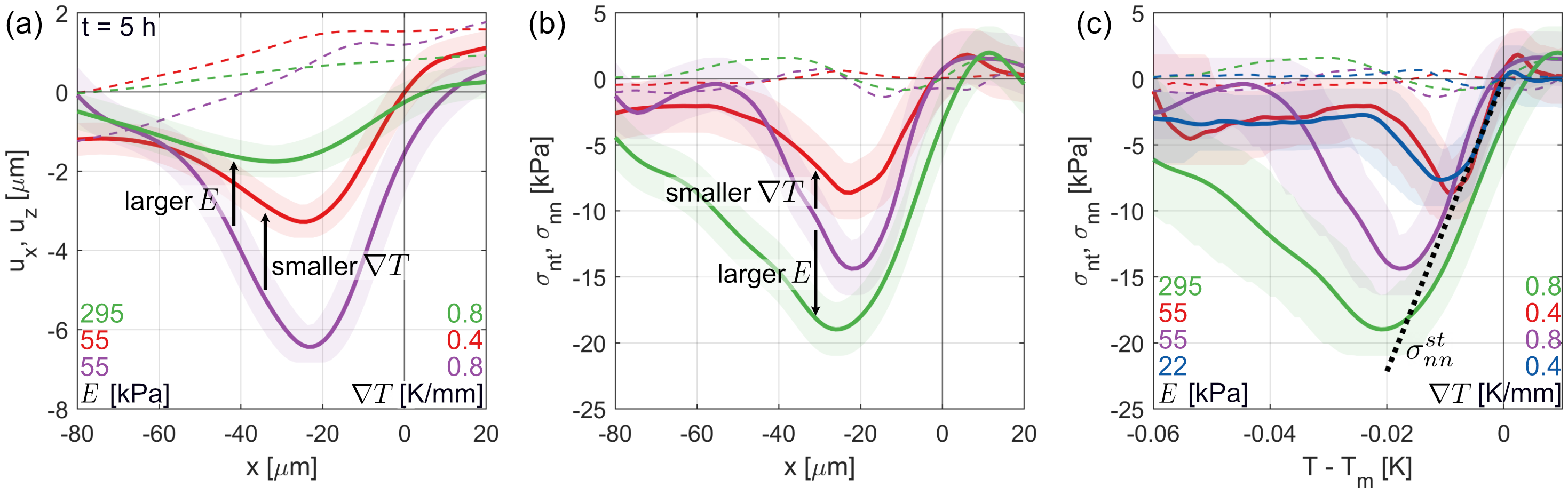}
    \caption{The effect of $E$ and $\nabla T$ on ice growth and stress build up
    (a) Substrate displacements $u_x$ (dashed curves) and $u_z$ (continuous curves) for four experiments with different applied temperature gradients and substrate stiffnesses, as indicated. The ice/water interface is at $x=0$.  (b) The corresponding stresses, $\sigma_{nt}$ (dashed curves) and $\sigma_{nn}$ (continuous curves). The shaded areas denote experimental uncertainties. (c) shows the same stresses as a function of the local temperature. The dashed line shows the expected stalling stress from equation~(\ref{eqn:clausius}). \emph{In all panels,} soft elastic layer thicknesses were \mum{88}, \mum{105}, \mum{105} and \mum{88}, while crystal orientations were $\theta= 30\pm 10^\circ$, $82\pm 8^\circ$, $82\pm 8^\circ$ and $82\pm 8^\circ$ for green, red, purple and blue curves respectively. The blue curves corresponds to the data in Figure~\ref{fig:deform_and_stresses_over_time}.}
    \label{fig:compare_d_and_s}
\end{figure*}

We see similar results when we vary $\nabla T$ and $E$.
Figure~\ref{fig:compare_d_and_s}a shows the indentation at the same timepoint for three experiments with similar $h$.
Increasing stiffness results in smaller indentations (compare green and purple curves), while increasing the temperature gradient results in a faster ice build-up (compare red and purple curves).
Figure \ref{fig:compare_d_and_s}b shows the calculated stresses corresponding to the data in Figure \ref{fig:compare_d_and_s}a .
As before, there are always negligible shear stresses, and the normal stresses locally stall near the ice-water interface (see Supplement).
Interestingly, while the size of the indentation is very sensitive to substrate stiffness, the stresses are much less so.
Here, the stresses near the ice/water interface are very comparable for the two experiments with the same temperature gradient, despite a factor of six in stiffness.
On the other hand, increasing the temperature gradient by a factor of two seems to approximately double $\sigma_{nn}$ near the ice-water interface.
These results hint at a stall pressure that mainly depends on local temperature.

A local, temperature-dependent stall pressure fits remarkably well with the predictions of Eqn. \ref{eqn:clausius}.
Stalling should occur when flow from the bulk water into the premelted layer ceases: when $P_l=0$.
This corresponds to a predicted stall stress of $\sigma^{st}_{nn}=-\rho q_m\frac{T_m-T}{T_m}$.
In Figure~\ref{fig:compare_d_and_s}c, we plot the measured stresses as a function of $T$, along with the theoretical prediction (dashed line).
We see that the measured stall stresses collapse onto a single line that match the theory to within error bars.
Interestingly, this collapse is apparently not affected by the ice crystal orientation, despite the fact that different ice facets are known to have different premelting characteristics~\cite{furukawa_ellipsometric_1987,slater_surface_2019,dosch_glancing-angle_1995}.
This can be seen by comparing the green and purple curves, which have very different crystallographic orientations ($\theta=30\pm 10^\circ$, $82\pm 8^\circ$ respectively).
It is also intriguing that the theory works so well, as it relies on the assumption of isotropic stresses in the ice.
In our experiments this is certainly not true, as we measure large stress gradients along the ice interface.

While the details of the final, local, stall pressure only depend on $\nabla T$, the global growth dynamics are more complex.
Figure \ref{fig:compare_d_and_s}a shows that displacements build up fastest in higher temperature gradients, and on softer substrates.
By contrast, Figure \ref{fig:compare_d_and_s}b shows that stresses appear to build up fastest in steeper temperature gradients but on \emph{stiffer} substrates.
The distinction between these is important, as it is the stresses that dictate when damage is likely to occur.
Ultimately, the dynamics are controlled by the interplay between the pressure gradients in the premelted film, and how this film thins at lower temperatures.
This has been modelled for the case of a simple, spring-like substrate \cite{wettlaufer_dynamics_1995,wettlaufer_premelting_2006}, and our results are in qualitative agreement with that work.

In conclusion, we have characterized how stresses build up around an ice crystal in a steady temperature gradient.
Our technique allows us to measure local stresses with $\mathcal{O}(\mu m)$ resolution.
Near the ice/water front, ice grows by cryosuction, causing normal stresses of $\mathcal{O}(\mathrm{kPa})$ to build up within a few minutes.
These are extremely localized, explaining their large propensity for causing damage.
At the same time, ice exerts much smaller shear stresses on its surroundings, presumably because of the lubricating effect of premelted layers between the ice and substrate.
Ultimately, the normal stresses reach a stall value, of about 1 MPa/K, which is in remarkably good agreement with the long-standing predictions of the Clapeyron equation, despite its inherent simplifications.
This gives strong support to this equation's widespread use as a foundation of freezing theory, and in other systems where premelting is important, including glacier movements~\cite{rempel_premelting_2019}, and alloys and metals at high temperatures \cite{nenow_surface_1989, hickman_disjoining_2016}.
Furthermore, our work is closely related to the crystallization pressure observed when salt crystals form in confinement~\cite{sekine_situ_2011, rijniers_experimental_2005,desarnaud_pressure_2016}, and the condensation pressure in liquid-liquid phase separation \cite{rosowski2020elastic}.
In these cases,  pressures caused by supersaturation result in a stress build up, and our technique offers a simple way to study this.
Interestingly, our technique also provides a way to measure local disjoining pressures in such systems.
At the stall point, $\Pi=\sigma_{nn}$, so our measurements of the stresses are also a local measurement of $\Pi$, with $\mathcal{O}(\mu\mathrm{m})$ resolution.

There is clearly still work to be done to understand the cryosuction process.
Here it takes a relatively long time to build up micron-sized deformations with accompanying stresses of $\mathcal{O}(10\mathrm{kPa})$. 
However, in practice, materials can fail even after one short freezing cycle and with smaller temperature gradients.
This indicates that faster transport mechanisms are likely present than the premelted-film flows that dominate here.
One potential mechanism is the flow of supercooled water through porous substrates like hydrogels or biological tissue.
Alternatively, water normally contains significant amounts of solutes, and this is known to strongly influence both bulk freezing \cite{worster_convection_1997}, and premelted layer characteristics \cite{wettlaufer_impurity_1999}.
We hope to address this question in future work.

This research was supported by ETH Research Grant ETH-38~18-2. 
We thank John Wettlaufer and Ian Griffiths for helpful discussions.

There are no conflicts of interest to declare.




\setcounter{figure}{0}
\renewcommand{\thefigure}{S\arabic{figure}}

\clearpage
\section*{Supporting Information}

\subsection{Temperature controller and Microscopy}
\begin{figure*}[htbp]
    \centering
    \includegraphics[width=\linewidth]{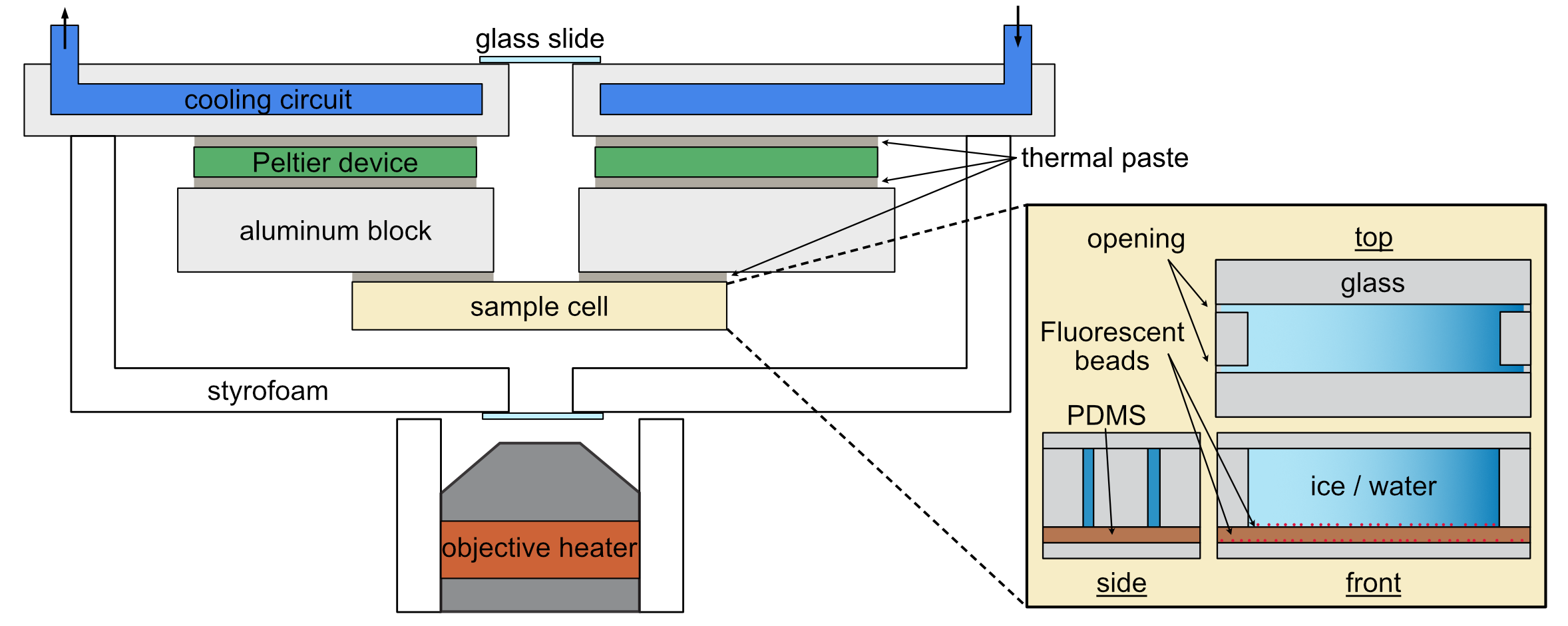}
    \caption{Schematic of the experimental setup (see text for dimensions). }
    \label{fig:setup_schematic}
\end{figure*}

Ice made from ultra-pure water was studied in cell with a stable temperature gradient over time periods up to 5~hours.
The cell is attached with thermal paste to two aluminium blocks, as shown in Figure~\ref{fig:setup_schematic}.
The blocks contained a thermistor, and were mounted under peltier elements.
We could then precisely control the temperature of the individual aluminium blocks by using a microcontroller with a feedback loop connected to the thermistors and peltier elements.
The excess heat generated by the peltier elements was removed with a cooling circuit that was attached to a heating/cooling circulator bath (Julabo F12-MA) that was kept at 23~°C. 
To minimize temperature fluctuations, the whole set up was insulated with polystyrene foam.
With this setup, the position of the ice/water interface fluctuated less than \mum{10} over the course of the experiments, lasting up to 5~hours. 

The freezing setup fits into the piezo stage (ASI PZ-2300) of a confocal microscope (Intelligent Imaging Innovations).
The objective used is a Nikon 20x air objective (NA~0.45, working distance 6.9-8.2~mm). 
The resulting field of view is 674 x \mum{674} recorded on 2048~x~2048 pixels.
To stabilize the microscope objective's temperature it was heated by a heating collar (Bioscience tools TC-HLS-025) to 27.5~°C throughout the experiment.

\subsection{Experimental cell} 
The experimental cell was made from standard microscope slides and UV-curable glue (NOA-61).
Microscope slides were used as spacers to create interior dimensions of approximately 3~x~1.5~x~0.6~mm.
At the left and right side of the cell, holes were left in the spacer, to prevent the build up of pressure due to the expansion of water as it freezes as can be seen in Figure~\ref{fig:setup_schematic}. 
In order to quantify the pressures onto the cell walls, the lower side of the experimental cell consisted of a soft elastic layer spin-coated onto a glass slide. 
The soft elastic layer was made from silicone (Gelest \cite{style_liquid-liquid_2018} or Sylgard 184) with Young's modulus between 22 and \SI{255}{\kilo \pascal} and thickness from 85 to \mum{105}.
The deformation of this layer was measured by tracking fluorescent tracer particles (Thermo Fisher Scientific, carboxylate-modified, 200~nm, red) deposited at the top and bottom interface of the soft elastic layer. 
To deposit them onto the glass substrate the glass was first treated with Ozone (UV/Ozone ProCleaner Plus, Bioforce Nanosciences) and silanized with (3-Aminopropyl)triethoxysilane vapour. 
To deposit the particles on top of the silicone layer a particle dispersion was simply put on top of it for 30 minutes, removed and rinsed with water.
The fluorescent particle dispersion consisted of borate buffer (400 ml water, 2.88~g Sodium tetraborate decahydrate, 8.9g~boric acid, pH 7.4)/EDC (1-Ethyl-3-(3-dimethylaminopropyl)carbodiimide) (5~mg in 500~\textmu l)/fluorescent particles (200~nm diameter, carboxylate modified) in the ratio of \emph{e.g.} \mbox{20~ml/500 \textmu l/14 \textmu l}.
With the deformations and the knowledge about the mechanical properties of the soft elastic layer the stresses exerted on the elastic layer could be calculated according to Traction Force Microscopy (TFM) \cite{xu_imaging_2010,style_traction_2014}.

\subsection{Growing Ice} 
The cells described above were filled with ultra-pure water with a very dilute concentration of fluorescent particles (Thermo Fisher Scientific, carboxylate-modified, 100~nm, yellow-green, 4~\textmu l in 20~ml) and mounted onto the temperature controller stage. 
The freezing point of the particle-containing water was measured with a freezing point osmometer (Gonotec Osmomat 3000) to ensure that the particles do not change the freezing behaviour of the water.
The particles dispersion was stored at room temperature, meaning that it was undersaturated with air when cooled to the experimental conditions, which helped to prevent the formation of air bubbles in the water phase.
The temperatures at the two Peltiers were set just above and below the freezing temperature and the system was allowed to equilibrate for 5 minutes. 
Ice was then nucleated by touching the cold side of the experimental cell with a cotton swab that had been cooled in liquid nitrogen.
The temperature was then lowered slowly ($-0.1$~K/min) to grow the ice crystal into the space between the two Peltier elements where it could be observed on the microscope. 
The slow growth is necessary to produce large single crystals of ice.

After reaching a temperature profile close to the target the setup was allowed to equilibrate again for 15 minutes. 
After this, a measurement series was started, with the temperature being lowered to the target setting as the first image was acquired.

\subsection{Estimation of the crystal orientation}
\begin{figure}[htbp]
    \centering
    \includegraphics[width=\linewidth]{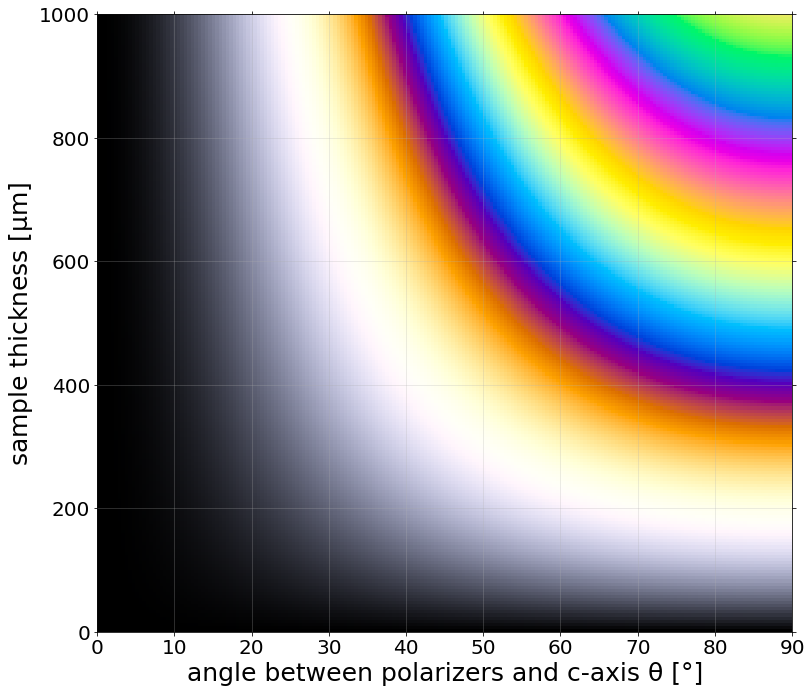}
    \caption{Adapted Michel-Lévy interference color chart showing the expected interference color of a ice single crystal in dependence of its thickness and orientation angle $\theta$. Generated with the help of \cite{m_bauer_githubcalculated_michel_levy_chart_2019, sorensen_revised_2013}. }
    \label{fig:gerber_chart}
\end{figure}

The ice crystal orientation was estimated by observing its color in crossed polarizers.
The interference color of an ice crystal in crossed polarizers only depends on its thickness and the orientation of the crystal with respect to the axis between the polarizers~\cite{bloss_optical_1999}.
Since we can measure the thickness of the ice crystal with the confocal microscope, we can directly relate the observed color to the ice crystals orientation.
For practical use we created an adapted Michel-Lévy interference color chart (Fig.~\ref{fig:gerber_chart}) of the expected interference color in relation to the ice thickness and orientation with the help of code by Bauer~\cite{m_bauer_githubcalculated_michel_levy_chart_2019} and analogous to Sørensen~\cite{sorensen_revised_2013}.
The dependence of the birefringence $\Delta n$ on $\theta$
 \begin{equation}
   \Delta n(\theta) =  n_e n_o / \sqrt{n_e^2 cos^2\theta + n_0^2sin^2\theta} - n_o,
   \label{eqn:birefringence_theta}
\end{equation}
was taken from \cite{wilen_development_2003}, where $n_e$ and $n_o$ denote the extraordinary and ordinary indices of refraction at normal incidence (1.3104 and 1.3090 respectively, at 590~nm).
The color on the map is falsified by the conversion to RGB and printing with arbitrary printers or displays, however we expect a precision of approximately 10°. 
We note that this method could be improved by using a spectrometer and fitting the measured color to the prediction from theory.

To use this chart, it is imperative to make sure that the observed crystal is a single crystal.
This can be checked by rotating the polarizer, while keeping the analyzer fixed and imaging the ice-water interface.
When light coming through the polarizer is aligned or perpendicular to  the optical axis of an ice crystal, it will travel through the ice unaffected by its birefringence.
At this angle, light travelling through water will be identical to light that has travelled through a single ice crystal.
At other angles, light travelling through the two materials and then through the analyzer will appear different.
Thus, if we have a single ice crystal adjacent to water there will only be four polarizer angles where the ice and water have identical color as the polarizer.
If there are multiple ice crystals stacked on top of each other in the cell, there is generally no angle where the light is unaffected by combination of the birefringences of the ice crystal layers.
Thus, one will not find any angles where water and ice appear identical as the polarizer is rotated.
This simple test makes it easier to identify single crystals.

\subsection{Measurement of mechanical properties}
Macroscopic samples of the used silicone were fabricated from the same mixtures that were used for spin coating.
Approximately 20~g of mixed silicone were cured in glass Petri dishes (height  \SI{1}{\centi \metre}, diameter \SI{5.5}{\centi \metre}) and cured under the same conditions as the spin-coated TFM substrates.
Mechanical testing was done with a texture analyzer (Stable Microsystems) with a cylindrical indentor (diameter 2~mm).
The force response of indentation was analyzed with MATLAB to calculate the Young's Modulus of the bulk silicone. 
A correction for the thickness of the sample was applied after \cite{garcia_determination_2018}.

\subsection{Temperature gradient measurement}
Temperatures in the cell were measured with thermistors (Adafruit B3950 NTC) that have been calibrated with a high precision thermal bath (Julabo F12-MA) and an ice bath.
The resulting accuracy is 0.05~°C.
The thermistor was placed in the experimental cell at the edge of the warm Peltier element towards the center of the thermal stage.
The temperatures in the cell were then measured under the experimental conditions on the microscope. 
By measuring this temperature and the distance of the thermistor to the ice/water interface the temperature gradient in the water could be calculated. 
The temperature gradient in the ice was derived by taking into account the ratio of the thermal conductivities of ice and water.

\subsection{Fluorescent particle localization}
After the acquisition of the confocal data the tracer particles on the substrate surface were located with a bespoke particle tracking code in MATLAB. 
The images of individual particles exhibited a large point spread function because of the low numerical aperture objective that was used.
To determine the z-position precisely we plotted the maximum intensity pixel for each image of a single particle (i.e. vs $z$) and fitted a Gaussian function through this.
The peak was then taken as the particle's sub-pixel$z$-position.
For the ($x$,$y$) position the weighted centroid of the particle image in the image taken closest to the sub-pixel $z$-position was used.
Because we image the particles deposited at the silicone/water interface through silicone, we corrected their $z$ coordinates by $z_{real} = z_{measured} \frac{n_{sil}}{n_{air}}$, where $n$ is the refractive index of the material~\cite{hell_aberrations_1993,besseling_methods_2015}.

Particles were localized on the silicone/water surface but also on the silicone/glass surface.
This allowed us to correct for the radial distortion of the imaging system (flat surfaces appear slightly parabola-shaped) by using the silicone/glass interface as a reference flat surface.

\subsection{Data treatment for traction force microscopy (TFM)}

\begin{figure}[htbp]
    \centering
    \includegraphics[width=\linewidth]{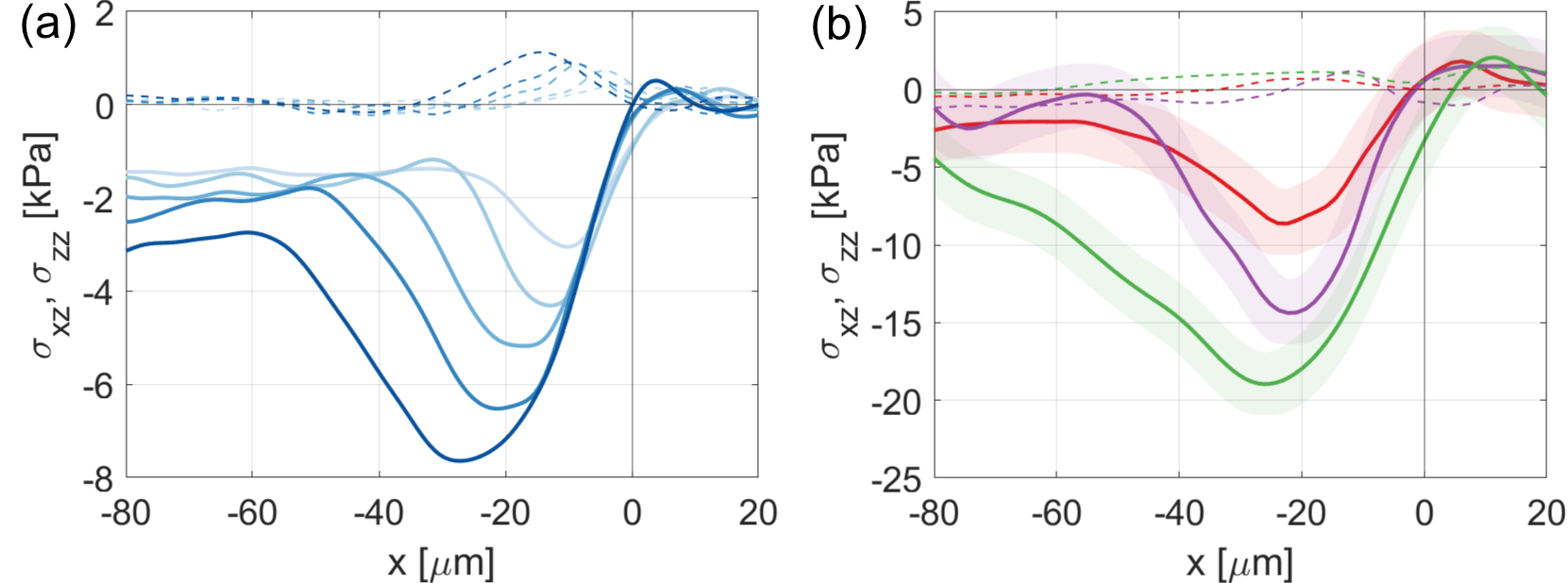}
    \caption{Stresses  $\sigma_{xz}$ (dashed line) and $\sigma_{zz}$ (continuous line) corresponding to (a) Figure~\ref{fig:deform_and_stresses_over_time}b and (b) Figure~\ref{fig:compare_d_and_s}b. }
    \label{fig:supplements_xz_and_zz}
\end{figure}

Displacements were calculated by connecting particle location in a deformed state to their position in a reference state -- just before the last advancement of the ice front.
This data was then collapsed parallel to the ice/water interface to obtain a collapsed data set of displacements ($u_x$,$u_z$) as a function of the distance to the ice/water contact line on the substrate.
These displacements where then interpolated to produce equally spaced data points for the TFM analysis.
At each designated $x$-position, the 25 closest neighbouring points were fitted with a parabola and the value of the parabola at this point was taken for the $z$-value.
Fourier smoothing was applied to the displacement data to reduce the noise on the calculated stresses.
Optimal smoothing parameters were found in an iterative process by minimizing the ratio between the integral of the stresses in the water part of the cell (supposed to be 0) by the integral of the stresses (supposed to be non-zero) under the ice.
The ratio of the integrals always had a clear minimum, which was around a low-pass smoothing parameter of \mum{25}.
After this, TFM analysis was conducted as described in the main text. 
After calculation of $\sigma_{xz}$ and $\sigma_{zz}$, we calculated local normal and traction stresses $\sigma_{nt}$ and $\sigma_{nn}$ by a coordinate transformation.
Specifically, we calculated local normal and tangent vectors to the substrate surface, $\mathbf{n}$ and $\mathbf{t}$.
Then,
\begin{equation}
\sigma_{nt}=\mathbf{t}.\left( \begin{array}{c}
\sigma_{xz} \\
\sigma_{zz} \end{array} \right) \,\,\, \mathrm{and}\,\,\, \sigma_{nn}=\mathbf{n}.\left( \begin{array}{c}
\sigma_{xz} \\
\sigma_{zz} \end{array} \right)
\end{equation}
The resulting values of $\sigma_{nn}$ are almost identical to stresses in the z-direction $\sigma_{zz}$.
However, there is a small, but significant differences between the profiles for $\sigma_{nt}$ and $\sigma_{xz}$.
Plots showing the corresponding $\sigma_{xz}$ and $\sigma_{zz}$ stresses to Figure~\ref{fig:deform_and_stresses_over_time} and \ref{fig:compare_d_and_s} are shown in Figure~\ref{fig:supplements_xz_and_zz}

The position of the ice/water interface was determined with the signal from the fluorescent particles in the water phase (these are repelled by the ice as it grows).
We took a maximum intensity projection in the $z$-direction of the confocal image stack of the particles in the water.
This resulted in a 2D top-view image of the system with a sharp contrast between the water and ice phases at the position of the ice/water contact line on the substrate.
This data was fitted to determine the shape and position of the ice/water interface.
For subsequent analysis, the image and confocal data are rotated digitally so that the ice water interface is aligned exactly with the $y$-direction.

\subsection{Supplementary data}
\begin{figure}[htbp]
    \centering
    \includegraphics[width=\linewidth]{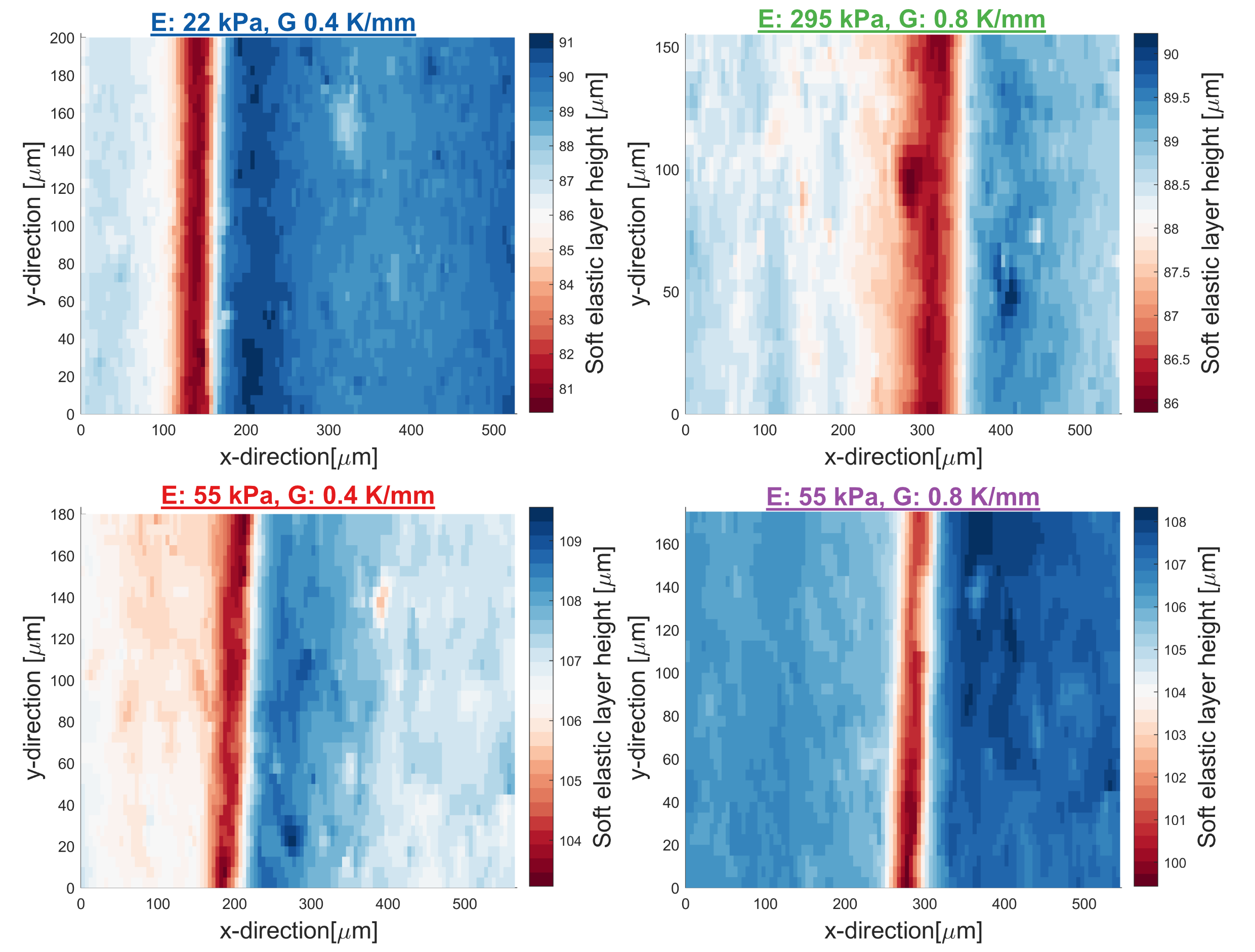}
    \caption{Surface profile of the soft elastic layer at the end of the experiments shown in Figure~\ref{fig:compare_d_and_s}. }
    \label{fig:all_topviews}

    \centering
    \includegraphics[width=\linewidth]{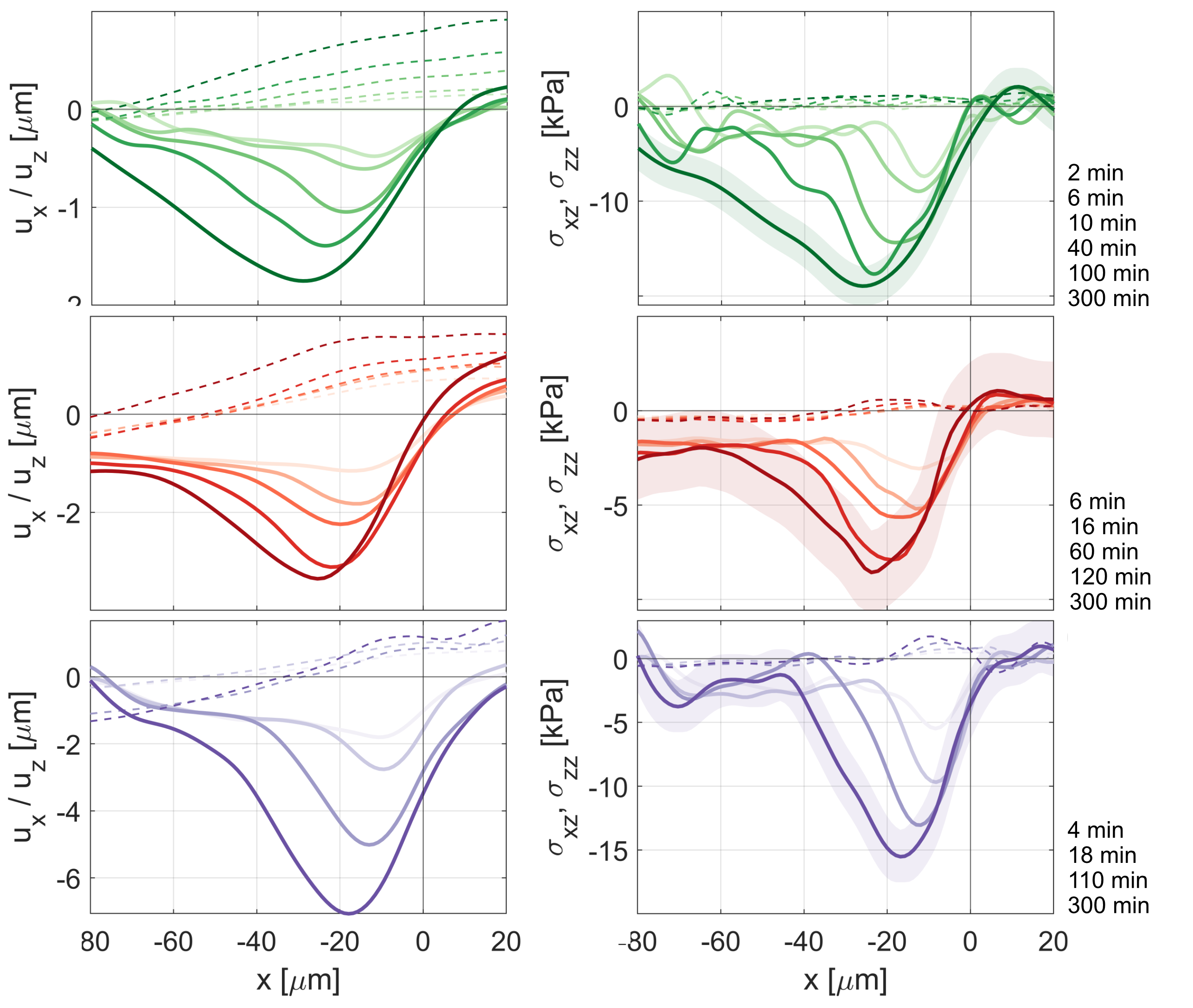}
    \caption{Time evolutions of the displacement and stresses of the experiments shown in Figure~\ref{fig:compare_d_and_s}. The time evolution of the blue curve is shown in Figure~\ref{fig:deform_and_stresses_over_time} of the main text. All curves are relative to the ice/water interface at the given timepoint. The uncertainty given by shaded area is representative for all curves and comes mainly from the fluctuations in ice/water interface position. }
    \label{fig:all_raw_data}
\end{figure}

The 3d data and time evolution corresponding to Figure~\ref{fig:compare_d_and_s} can be found in Figure~\ref{fig:all_topviews} and~\ref{fig:all_raw_data}. 

\begin{figure}[htbp]
    \centering
    \includegraphics[width=\linewidth]{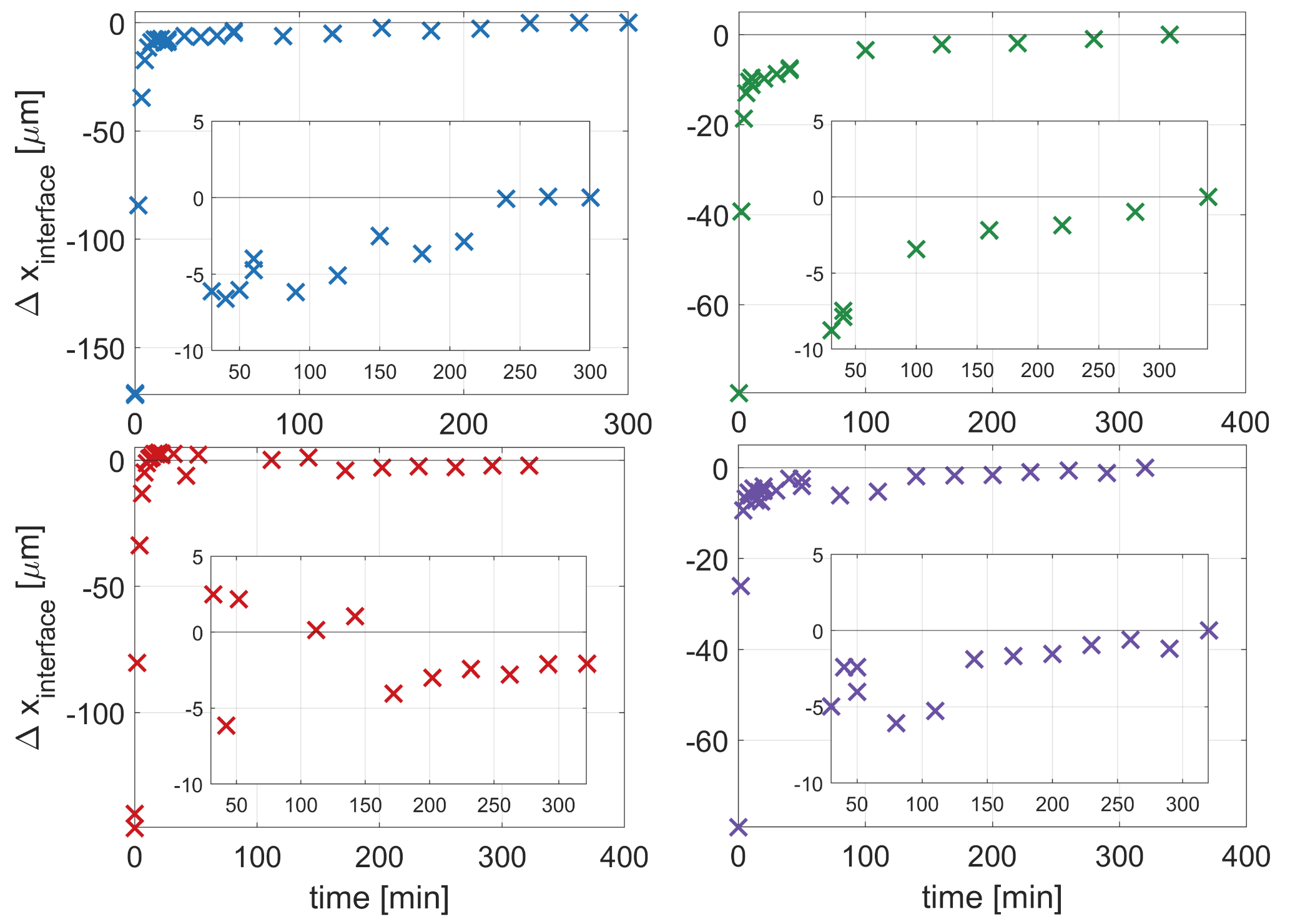}
    \caption{Ice/water interface position in relation to the last timepoint for the experiments shown in Figure~\ref{fig:compare_d_and_s}. The inset shows a zoomed in section of the same data. }
    \label{fig:interface_over_time}
\end{figure}

Figure~\ref{fig:interface_over_time} shows the movement of the ice/water interface over time. 
Generally the interface advances quickly after the last temperature step (applied at the start of the temperature) and very slowly during the rest of the experiment.
The undesired second advancement is likely to come from fluctuations in the room temperature.
Although undesired, these fluctuations of the ice/water interface position are not problematic, as the slope of the pressure next to the pressure curve next to the ice/water interface ($x=0$) adapts faster, than the interface moves.
This can be seen in Figure~\ref{fig:all_raw_data}, where the stresses are plotted in relation to the interface position at the timepoint of imaging.
The interface movement however limits the precision with which we can study the dynamics of the system, therefore only qualitative statements could be made.

\begin{figure}[htbp]
    \centering
    \includegraphics[width=\linewidth]{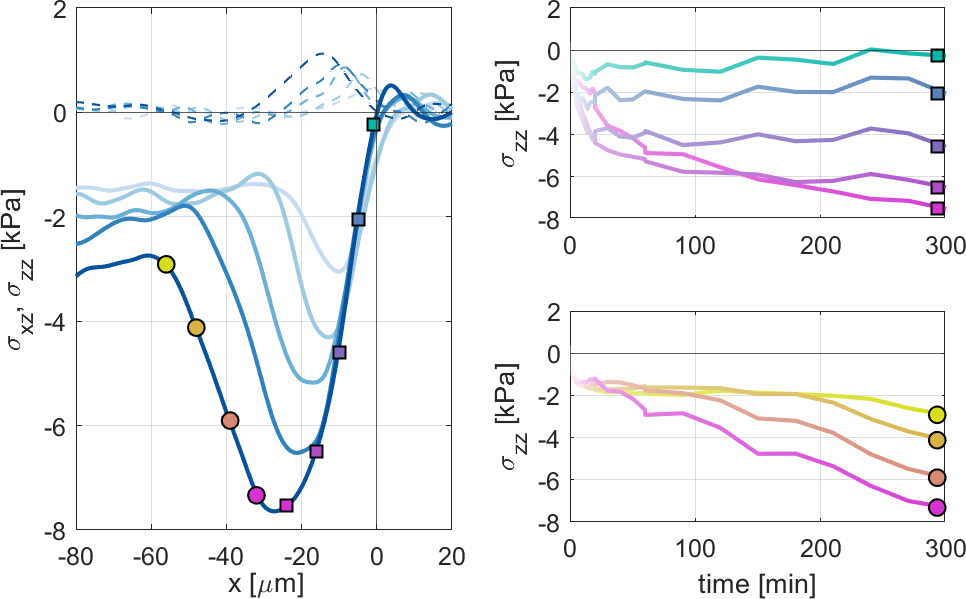}
    \caption{Extension of Figure~\ref{fig:deform_and_stresses_over_time} in the main text, where more time evolutions are shown for locations where the stall pressure is not yet reached (circle markers). }
    \label{fig:more_over_time}
\end{figure}

Figure~\ref{fig:more_over_time} shows additional stress evolutions at locations where the stalling pressure has not been reached.

\subsection{Error estimation}

The error in the displacement data stems from the imprecision with which the fluorescent particles can be located.
The fluorescent particles have a large point spread function, because of the low numerical aperture (NA 0.45) of the used objective, which makes them difficult to locate precisely. 
The precision in the $z$-direction of the locating method described above was tested by imaging a flat sample (fluorescent particles attached to a glass microscope slide), locating the particles and measuring the noise in the data relative to a plane fitted through all the particle positions.
The standard deviation of the noise was \mum{0.2}.
To measure the precision in the $x$ and $y$-direction the same particles were imaged twice (where the objective was moved out of focus in between) and the deviation of the positions in the two images was measured.
The standard deviation of this noise was \mum{0.03}.
For displacements, the error is doubled, as the errors at two different time points add up.

The reported stresses have three sources of error: the TFM algorithm, the displacement measurement and the imprecision of the mechanical properties of the soft elastic layer.
The TFM algorithm was checked with synthetic data (calculated displacements from known stresses with noise) to have a deviation of about 2~kPa.
The displacement error (discussed above) and the error in the measured Youngs' modulus (3~kPa) directly propagate to the stresses, giving an approximate error of
 \begin{equation}
   \Delta \sigma = E\Delta d / h + \Delta E  d/h,
   \label{eqn:error_stresses}
\end{equation}
with $h$ the thickness of the silicone layer, $E$ the Youngs modulus of the substrate, $d$ the displacement and $\Delta$ indicating the error in the following variable.

\begin{table*}
\caption{Properties of the experiments shown in Figure~\ref{fig:compare_d_and_s}}
\label{tab:properties}
\begin{tabularx}{0.8\textwidth}{c|c|c|c|c}
\textbf{line color} & \textbf{Young's modulus} & \textbf{temperature gradient} & \textbf{substrate thickness } & \textbf{Crystal orientation} \\ 
 & \textbf{$E$ [kPa]} & \textbf{$\nabla T$ [K/mm]} & \textbf{[$\mu$m]} & \textbf{[°]} \\ 
\hline
blue           &      22      $\pm 3$        &         0.4      $\pm 0.08$           &        88  $\pm 0.2$            &             82 $\pm 8$                 \\
red            &   55    $\pm 3$     &     0.4        $\pm 0.08$      &       105   $\pm 0.2$        &              82 $\pm 8$                \\
purple         &      55     $\pm 3$      &            0.8      $\pm 0.15$      &           105  $\pm 0.2$  &                  82 $\pm 8$            \\
green          &      295   $\pm 3$      &            0.8      $\pm 0.15$    &          88 $\pm 0.2$   &              30  $\pm 10$              
\end{tabularx}
\end{table*}

The error in the position of the ice/water interface comes from the imprecision of the fitting process and the slight tilt of the interface along the z-direction (see Figure~\ref{fig:setup}). 
The tilt is caused by the small temperature gradient in the $z$-direction from the relative warmth of the objective below the sample.
This blurs the line between water and ice in the maximum projection of the fluorescent particles dispersed in the water along the $z$-direction.
We estimate this error to be on the order of \mum{5}.

The reported temperatures in Figure~\ref{fig:compare_d_and_s}c have an error stemming from the temperature gradient measurement (described above).
The temperature gradient was calculated by measuring the temperature of the thermistor bead and its distance to the ice/water interface (where we know the temperature to be exactly 0°C).
This is prone to an error, because of the low precision of the temperature reading (0.1°C) and the relatively large size of the thermistor bead (\mum{200}).
It is assumed that the temperature gradient from the ice/water interface to the thermistor is linear, which is supported by a COMSOL simulation of our experimental setup.
We note here that the presence of the thermistor itself could influence the local temperature gradient (as its thermal conductivity is different from the water it replaces in the cell).
However, we estimate this effect to be negligible compared to the other sources of error, and this is supported by the fact that the ice/water interface position did not visibly deviate near the thermistor.
The ice/water interface was taken as the 0~°C isotherm, from which the temperature in the cell was extrapolated with the calculated temperature gradient to give the local temperature. 
The uncertainty in temperature thus increases with the distance to the ice/water interface.

The shaded area in Figures~\ref{fig:compare_d_and_s} and \ref{fig:all_raw_data} are a combination of all the error sources in the respective experiments.

\end{document}